\documentclass[12pt, draftclsnofoot, onecolumn]{IEEEtran}

\usepackage{cite}

\ifCLASSINFOpdf
  \usepackage[pdftex]{graphicx}
  \graphicspath{{./}{../pdf/}{../jpeg/}}
  \DeclareGraphicsExtensions{.pdf,.jpeg,.png}
\else
  \usepackage[dvips]{graphicx}
  \graphicspath{{../eps/}}
  \DeclareGraphicsExtensions{.eps}
\fi
\usepackage{amsmath,amssymb}
\usepackage{algorithmic}

\usepackage{array}

\ifCLASSOPTIONcompsoc
 \usepackage[caption=false,font=normalsize,labelfont=sf,textfont=sf]{subfig}
\else
 \usepackage[caption=false,font=footnotesize]{subfig}
\fi
\usepackage{url}
\usepackage{color,soul}
\usepackage{bm}
\usepackage{amsfonts}
\usepackage{algorithm}
\usepackage{caption}
\usepackage{booktabs}
\usepackage[normal]{threeparttable}
\usepackage{multirow}
\usepackage{cases}
\usepackage{diagbox}
\usepackage{colortbl}
\definecolor{mygray0}{gray}{1}
\definecolor{mygray1}{gray}{.90}
\definecolor{mygray2}{gray}{.85}
\definecolor{mygray3}{gray}{.75}
\definecolor{mygray4}{gray}{.65}
\definecolor{mygray5}{gray}{.55}
\definecolor{mygray6}{gray}{.45}
\definecolor{mygray7}{gray}{.35}
\definecolor{mygray8}{gray}{.25}
\definecolor{mygray9}{gray}{.15}
\definecolor{mypink}{rgb}{.99,.91,.95}
\definecolor{mycyan}{cmyk}{.3,0,0,0}

\begin{document}

\title{Indoor Localization Using Smartphone Magnetic with Multi-Scale TCN and LSTM}


\author{Mingyang~Zhang,~Jie~Jia,~Jian~Chen
\thanks{M. Zhang, J. Jia and J. Chen are with the School of Computer Science and Engineering, Northeastern University, Shenyang 110819, China (e-mail: \{zhangmingyang, jiajie, chenjian\}@mail.neu.edu.cn)}
}

\maketitle
\vspace{-2cm}
\begin{abstract}
   A novel multi-scale temporal convolutional network (TCN) and long short-term memory network (LSTM) based magnetic localization approach is proposed. To enhance the discernibility of geomagnetic signals, the time-series preprocessing approach is constructed at first. Next, the TCN is invoked to expand the feature dimensions on the basis of keeping the time-series characteristics of LSTM model. Then, a multi-scale time-series layer is constructed with multiple TCNs of different dilation factors to address the problem of inconsistent time-series speed between localization model and mobile users. A stacking framework of multi-scale TCN and LSTM is eventually proposed for indoor magnetic localization. Experiment results demonstrate the effectiveness of the proposed algorithm in indoor localization.
   \end{abstract}
   

   %
   \IEEEpeerreviewmaketitle

   \section{Introduction}
   %
   %
   %
   %
   \IEEEPARstart{T}{he} unprecedented expansion of new indoor location-based services is expediting the development of indoor localization techniques~\cite{zhou2020app,kim2019workload}. Demand for precise and timely location-based services (LBS) is increasing. With the burgeoning development of portable sensing techniques, indoor localization based on smartphones is gaining importance in various applications. The smartphone-based indoor localization techniques utilize the sensors embedded into a smartphone to collect the signals of the localization environment and localize a smartphone. Geomagnetic localization is one of the most promising smartphone-based indoor localization techniques, since it can utilize the ubiquitous geomagnetic field, avoiding the deployment of specialized hardware. Hence geomagnetic localization techniques have received remarkable attention both in the world of academia and industry.
   
   In the indoor environment, the received signal strength (RSS) localization~\cite{bahl2000radar}, as a typical fingerprint-based indoor localization approach, can locate a smartphone with high accuracy, such as the wireless fidelity (WiFi) localization~\cite{husen2016indoor} and bluetooth low energy (BLE) localization~\cite{xiao20173dble}. However, the signals of WiFi and BLE are the scalar-based radio frequency (RF) signals, and the dimension of fingerprint signals can be expanded by adding wireless access points or BLE beacons. 
   For the geomagnetic localization, the geomagnetic signals are the vector-based environmental signals and the dimension of fingerprint signals are fixed 3. The lower dimension of geomagnetic signals results in the non-uniqueness of geomagnetic signals in the indoor environment. Therefore, fingerprint matching localization using instant geomagnetic data cannot achieve good performance due to the mismatching fingerprints. 
   Some researchers fuse magnetic with other localization techniques to improve the accuracy of magnetic localization. The magnetic data were used for heading direction in pedestrian dead-reckoning localization~\cite{relpf2016xie,emer2013zhang,imu2019poulose}. The magnetic localization was fused with PDR localization based on Kalman filter, particle filter, or hidden Markov model~\cite{magbase2019wang,shu2015magical,basmag2016ma}. However, the accuracy of these magnetic localization approaches has not been improved on their own. Furthermore, adjusting filter parameters of fusion localization requires vast experience. 
   To overcome the low discernibility problem of geomagnetic signals, the time series-based geomagnetic fingerprint matching approaches are introduced into magnetic localization, and a classic implement is the dynamic time warping (DTW) fingerprint matching localization~\cite{locateme2013subbu,shu2015magical}. However, the DTW-based localization approaches require time-consuming time series-based fingerprint matching, thus positioning pedestrians with large delay, especially in large-scale localization environments. 
   
   With the development of deep learning, many deep learning approaches have been applied in indoor localization successfully. 
   For example, in the RF-based localization, the indoor localization approaches with channel state information (CSI) of WiFi signals based on deep neural networks (DNN) were proposed by~\cite{csi2017wang,csi2019yin}. The indoor BLE localization approach based on deep autoencoder was proposed by~\cite{xiao20173dble}. 
   Since the deep learning approach can separate the localization process into training models off-line and predicting positions on-line, the deep learning-based localization approaches can predict a target in real-time, thus avoiding the time-consuming DTW-based fingerprint matching localization. Some magnetic localization approaches based on deep learning have been proposed. For example, a magnetic indoor localization approach using convolution neural networks (CNN) was proposed by~\cite{amid2018lee}. However, the study~\cite{amid2018lee} mainly focuses on the spatial context of magnetic signals in localization environments. The temporal context of magnetic signals with pedestrian's movement has not been considered into the localization model. In~\cite{bae2019largescale}, the temporal context of magnetic signals was utlized to localizing pedestrians, and the LSTM-based triaxial magnetic localization approaches was proposed, which used the 3-dimensional geomagnetic vector as the input of the LSTM model instead of the signal strength of magnetic vector. However, the feature dimension of LSTM is still relatively low. In~\cite{deepml2018wang}, distance matrix of magnetic was designed and used as the input of LSTM model to expand the feature dimension of LSTM. In~\cite{zhang2021lstm}, a double sliding-window approach was designed to expand the feature dimension of LSTM. However, these approaches expand the feature dimension of LSTM by preprocessing the magnetic data, and the LSTM models are still not suitable for low dimensional magnetic data. In addition, the problem of the inconsistent time-series speed between localization model and mobile users is still unresolved.
   
   This paper aims at exploiting a novel indoor magnetic localization algorithm based on multi-scale TCN and LSTM for a smartphone to address the above problems. Unlike the fingerprint matching localization approaches, the proposed algorithm does not need time-consuming time series-based fingerprint matching. The localization estimation based on neural networks is executed in real-time. Inspired by the potential prediction ability of LSTM, this paper utilizes the LSTM networks to estimate pedestrian's position. The TCN is invoked to expand the feature dimensions on the basis of keeping the time-series characteristics of LSTM model. By constructing the multi-scale time-series layer with multiple TCNs of different dilation factors, the problem of inconsistent time-series speed between localization model and mobile users are solved. A stacking framework of multi-scale TCN and LSTM is eventually proposed for indoor magnetic localization. 
   \begin{figure}[!t]
     \centering
     \includegraphics[width=14cm]{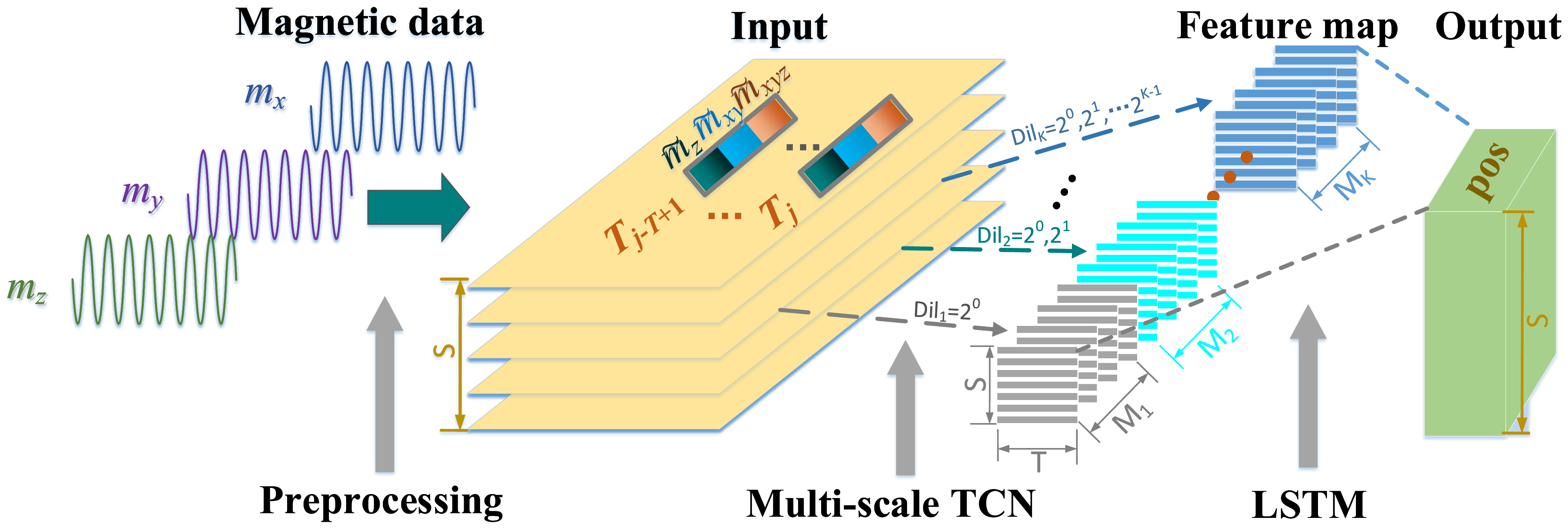}
     \caption{Framework of the proposed algorithm.}
     \label{fig:tcnlstmframework}
     \end{figure}
   
   \section{Proposed Algorithm}\label{sec:lstm}
   To present our proposed localization algorithm clearly, a framework of the indoor magnetic localization based on multi-scale TCN and LSTM is constructed as shown in Fig.~\ref{fig:tcnlstmframework}. 
   
   \subsection{Data Collection and Preprocessing}\label{sec:col&proc}
   The deep learning-based magnetic localization consists of two main phases: the off-line phase and the on-line phase. In the off-line phase, a site survey is conducted to collect magnetic signal features $\bm m=[m_x,m_y,m_z]$ at $Q$ known locations, the so-called reference points (RPs). With the data collected in RPs, the magnetic localization models are trained with deep learning methods. In the online phase, the user's position is estimated by invoking the trained deep learning models with the signal data of the target point.
   Since the proposed method localizes a target by utilizing the temporal context of magnetic signals, the collected magnetic samples need to be time-series. In indoor environments, the pedestrian's mobility is restricted by the environment, such as the walls, desks, cabinets, \textit{etc}. Therefore, not all possible movements within indoor environments are realized. Rather, the pedestrians often move along a limited set of typical trajectories~\cite{lee2008mobloc}. Therefore, we collect the magnetic data along the typical trajectories. 
   
   With the sampling data from the typical trajectories, the time-series characteristics of magnetic signals can be obtained by preprocessing the collected samples. The data preprocessing consists of the data normalization and data serialization. The data normalization is well studied and we employ the data normalization techniques used in~\cite{zhang2021lstm}. For the data serialization process, the normalized magnetic data $\bm m_t'=[\tilde{m}_z,\tilde{m}_{xy},\tilde{m}_{xyz}]$ are processed with one sliding-window method instead of the double sliding-window method used in~\cite{zhang2021lstm} or the distance matrix method in~\cite{deepml2018wang}, since the feature dimension of magnetic data do not need to be expanded in the data preprocessing of our method. The time-series magnetic feature can be represented as 
   \begin{equation}\label{eq:tsdim}
     \bm T_t=\begin{bmatrix}
             \bm m_{t-T+1}'\\
             \vdots\\
             \bm m_t'
             \end{bmatrix}
   \end{equation}
   where the $T$ is the sliding-window size. The $\bm T_t$ is the time-series magnetic features, which can be used as the input of deep learning model, and the corresponding postion $\bm X_t$ of RP can be used as the output label of deep learning model. When the process of data serialization is finished, the deep learning model can be trained with time-series magnetic features and the corresponding postion labels.
   
   \subsection{Multi-scale TCN}
   By preprocessing the magnetic data, the time-series dimension of magnetic data is created, however, the feature dimension of magnetic data is still 3. 
   In this section, the feature dimension of magnetic data will be expanded by invoking the TCN~\cite{BaiTCN2018}, a temporal variant of convolutional neural network (CNN). With the temporal context of magnetic data, various feature maps can be constructed to extract magnetic features by using the TCN, thus expanding the feature dimension on the basis of keeping the time-series dimension of magnetic data. 
   
   The feature dimension of magnetic data is expanded with the TCN, however, 
   when the speed of sampling magnetic data is inconsistent with that of positioning user, the deep learning model trained with the magnetic samples will not predict the user's position accurately. To tackle the problem, a multi-scale time-series layer is constructed with the TCNs of different dilation factors. Since the TCN employs the dilated convolutions to enable an exponentially large receptive field, the dilation factor is employed for constructing the TCNs with different receptive fields. The dilation factor can be written as
   \begin{equation}\label{eq:dil}
     \bm {Dil}_k=\{2^0,2^1,\cdots,2^{k-1}\}
   \end{equation}
   where the $\bm {Dil}_k$ is dilation factor, a set of exponential powers of 2. The $k$ is the number of the dilated convolution layers in the $k$-th TCN, and the dilation rate of the $k$-th dilated convolution layer is $2^{k-1}$. Since the TCN also inherits the characteristics of causal convolutions, the prediction of any time only depends on the input of current time and its previous time. Therefore, the receptive field of $k$-th dilated convolution layer with a kernal size of 2 is $2^k$. The maximum receptive field needs to cover all the data in the sliding window. Therefore, the sliding-window size $T$ can be set to be the maximum receptive field $2^k$. When the $k$ is determined, a series of TCNs with different dilation factors will be obtained, thus achieving the multi-scale TCN.
   
   \subsection{Stacking of TCN and LSTM}
   With the multi-scale TCN, we can obtain various feature maps as shown in Fig.~\ref{fig:tcnlstmframework}, where the $S$ represents the number of samples, and the $M_i$ is the number of feature maps output by the $i$-th TCN. Unlike the CNN used in image processing, the TCN in this paper uses 1-dimensional convolution to process the time-series magnetic data, so the feature map is 1-dimensional time-series data. For each feature map, and $T$ is the sliding-window size. As observed from Fig.~\ref{fig:tcnlstmframework}, we define the data at the $t$-th time of the output of the $s$-th sample on the $k$-th TCN as $\bm o_{s,t,k}$, then the input of the $s$-th sample on the LSTM can be represented as
   \begin{equation}
     \bm i_s^{LSTM}=\begin{bmatrix}
       \bm o_{s,1,1}&\cdots&\bm o_{s,1,K}\\
       \vdots&\ddots&\vdots\\
       \bm o_{s,T,1}&\cdots&\bm o_{s,T,K}\\
     \end{bmatrix}
   \end{equation}
   where the $K$ is the number of the TCNs, and the $T$ is the time-series dimension of feature maps. Then the feature dimension $D_f$ of feature maps can be written as
   \begin{equation}
     D_f=\sum\nolimits_{i=1}^K{M_i}
   \end{equation}
   If all the $M_i$ have the same value $M$, feature dimension of feature maps will be $D_f=K\times{M}$. The feature dimension of magnetic data is thus expanded by invoking the multi-scale TCN. The temporal characteristics of magnetic data are preserved well. Then, the LSTM is stacked behind the multi-scale TCN to take over the feature maps. The depth of the LSTM needs to be consistent with the sliding-window size. Therefore, we assign the depth of the LSTM to be $T$. The output of LSTM is the user's position. The LSTM is eventually stacked behind the multi-scale TCN. Fig.~\ref{fig:tcnlstmframework} shows the overall stacking framework of multi-scale TCN and LSTM.

   \section{Performance Evaluation}\label{sec:exp}
   In this section, we first introduce the experimental environment setup. Then the performance of the proposed algorithm is compared with some state-of-the-art magnetic localization approaches in the indoor experimental scenarios of different scales. Finally, we evaluate the performances of our proposed localization algorithm in the indoor experimental scenarios with different moving speeds.
   \subsection{Experimental Setup}
   \begin{figure}[!t]
     \centering
     \includegraphics[width=9cm]{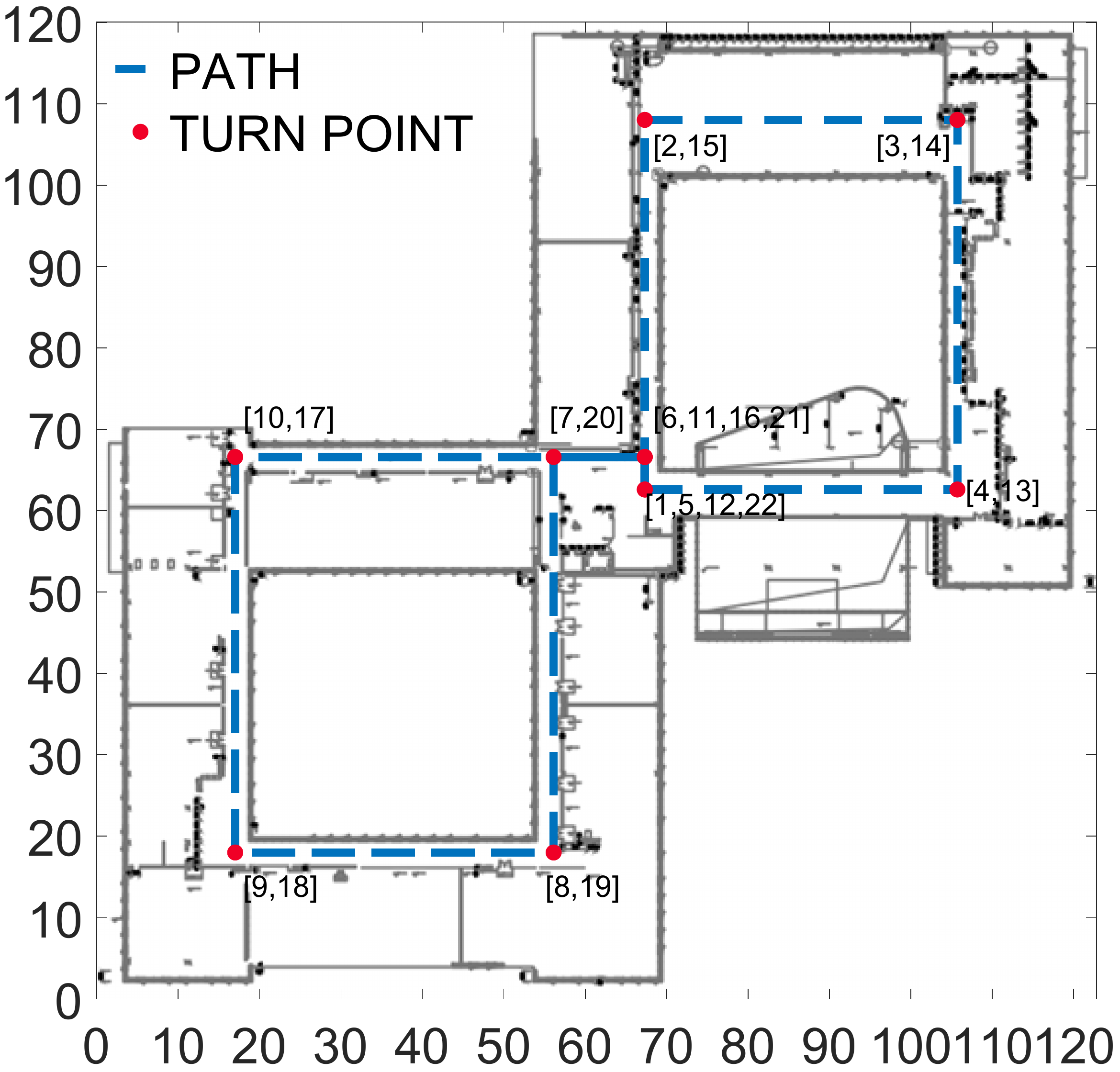}
     \caption{Views of experimental environment.}
     \label{fig:plottestbed}
     \end{figure}
   \newcommand{\RNum}[1]{\uppercase\expandafter{\romannumeral #1\relax}}
   \begin{table}[!t]
      
   \centering
   \begin{threeparttable}
   \caption{Specification of the experimental data}
   \label{tab:dataset}
   \small 
   \centering
   \begin{tabular}{cccc}
   \toprule
   \textbf{sceID}\tnote{a} & \textbf{Training Data} & \textbf{Validation Data} & \textbf{Testing Data}\\
   \midrule
   S&5215&1321&773\\
   M&11172&2927&1681\\
   L&21926&5814&3340\\
   \bottomrule
   \end{tabular}
   \begin{tablenotes}
   \item[a] The sceID corresponds to the three experimental scenarios: S. Small scale scenario. M. Medium scale scenario. L. Large scale scenario.
   \end{tablenotes}
   \end{threeparttable}
   \vskip -0.3cm
   \end{table}
   To evaluate the performance of the proposed algorithm in real environments, this paper gathers the geomagnetic sensor data of smartphones from the experimental environment as shown in Fig.~\ref{fig:plottestbed}. 
   The area of localization scenario is  122~m $\times$ 120~m. The sampling path contains 22 turning points. The mobile user completes the magnetic sampling once by walking from point 1 to point 22 along the path. This paper designs three scales of experimental scenarios: 1) small scale (points 1$\sim$5), 2) medium scale (points 1$\sim$11), and 3) large scale (points 1$\sim$22). The specification of experimental data is represented in TABLE~\ref{tab:dataset}. The SceID in TABLE~\ref{tab:dataset} corresponds to the three scenarios in Fig.~\ref{fig:plottestbed}. The numbers of training data, Validation data, and testing data for different scales are described in TABLE~\ref{tab:dataset} detailedly. The sampling frequency is set to 20Hz. For each scale, the sampling process is executed five times.
   The experiments in this paper are carried out on the Windows 10 64-bit system. CPU is Intel\textsuperscript{\circledR} Core\textsuperscript{TM} i7-9750H with a base frequency of 2.60 GHz. GPU is NVIDIA GEFORCE RTX2060 with CUDA 9.1 and cuDNN 7.2.1. The deep learning-based indoor magnetic localization task is deployed on the machine learning framework TensorFlow 2.3.0 with deep learning library Keras 2.4.3. To prevent overfitting, the early stopping mechanism of TensorFlow is introduced into our proposed localization approach. 
   The parameters of our proposed localization algorithm are selected by a series of experiments and analysis. The sliding-window sizes for the small scale scenario, medium scale scenario, and large scale scenario are 64, 128, and 128, respectively. The number of multi-scale TCNs for the small scale scenario, medium scale scenario, and large scale scenario are 7, 8, and 8, respectively. The number of feature maps for each TCN is 10. The number of hidden units is 200.

   \subsection{Results of the Proposed Algorithm}
   In this section, the 
   proposed multi-scale TCN and LSTM-based magnetic localization algorithm (MSTL) is compared with the state-of-the-art DTW fingerprint localization~\cite{locateme2013subbu} (DTW), the conventional LSTM-based localization~\cite{bae2019largescale} (LSTM), the TCN-based localization~\cite{BaiTCN2018} (TCN) in indoor localization scenarios of different scales and moving speeds. 
   \begin{figure}[!t]
     \centering
     \vspace{-0.3cm}
     \subfloat[Result of sceID S]{\includegraphics[height=6cm]{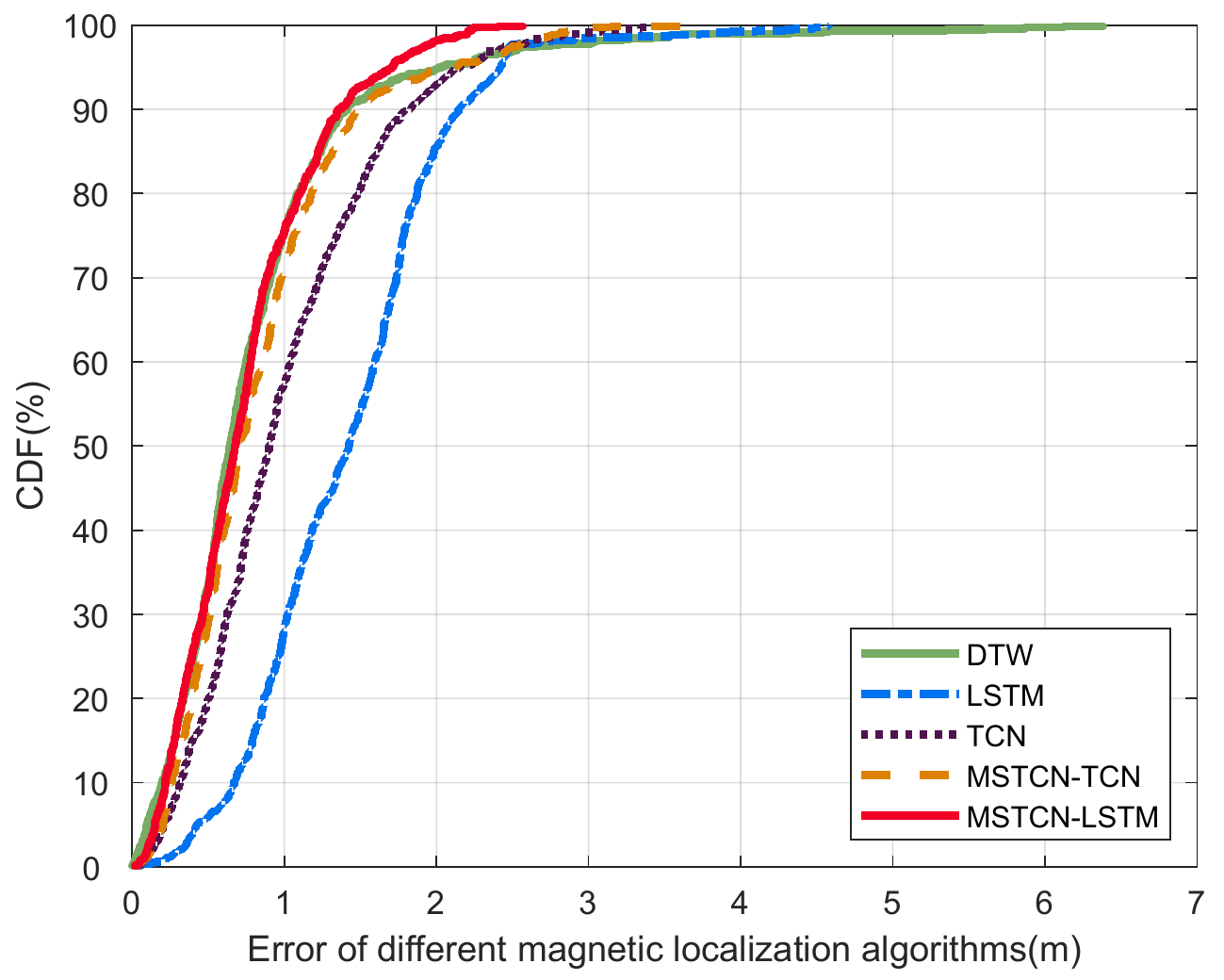}%
     \label{fig:cmpplotcdfareascalea}}
     \hfil
     \subfloat[Result of sceID M]{\includegraphics[height=6cm]{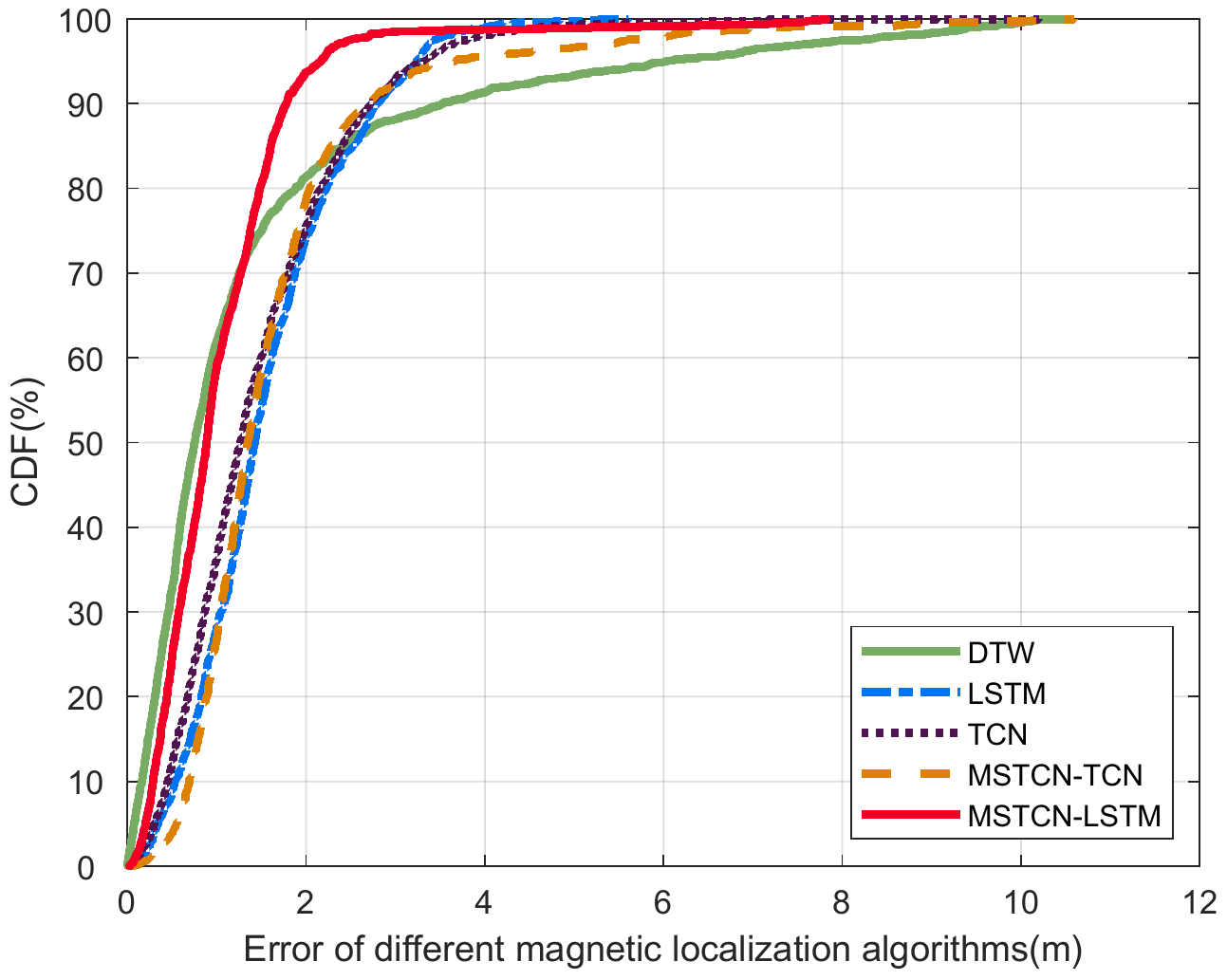}%
     \label{fig:cmpplotcdfareascaleb}}
     \hfil
     \subfloat[Result of sceID L]{\includegraphics[height=6cm]{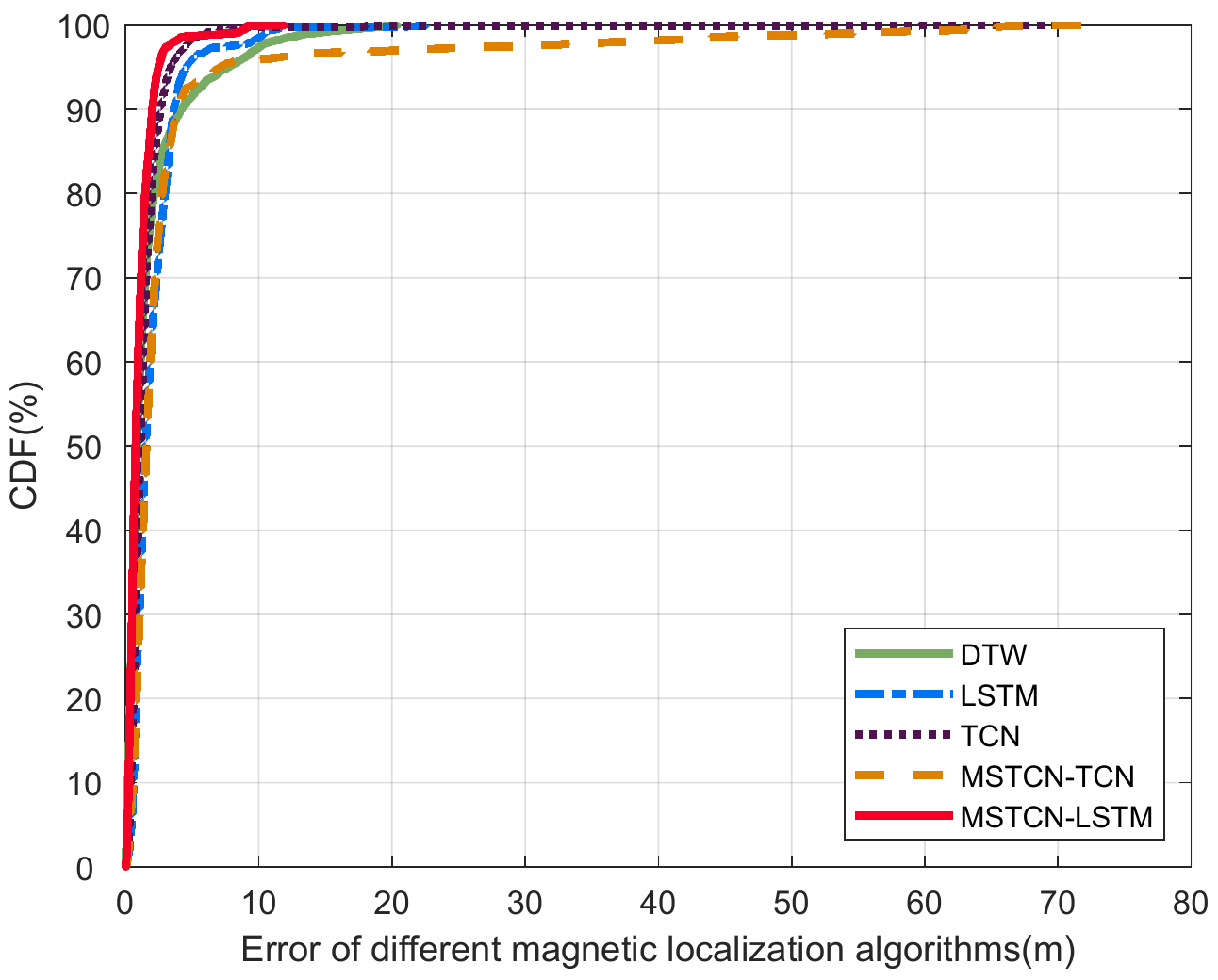}%
     \label{fig:cmpplotcdfareascalec}}
     \hfil
     \subfloat[Part of result of (c)]{\includegraphics[height=6cm]{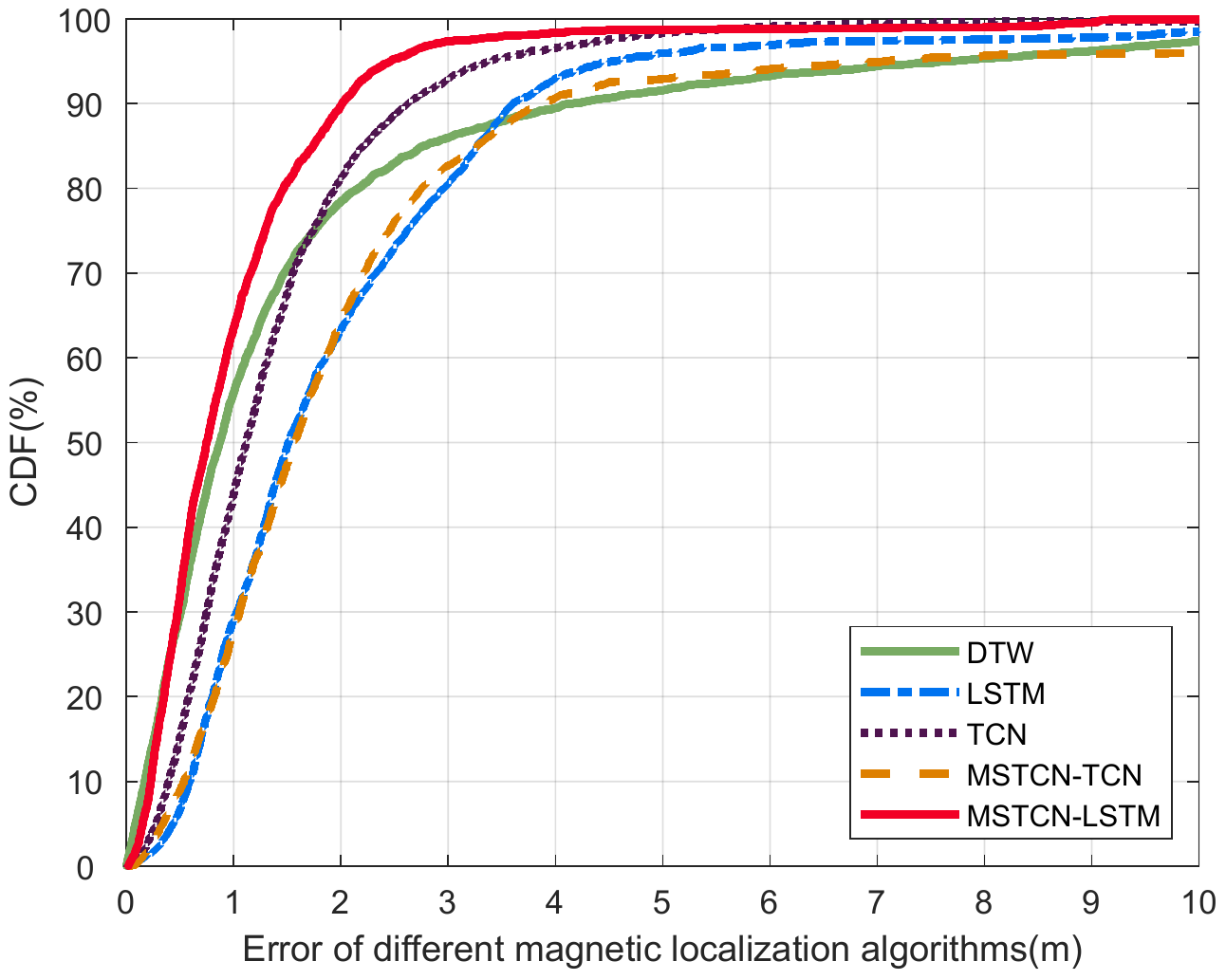}%
     \label{fig:cmpplotcdfareascaled}}
     \caption{Cumulative error distributions of different magnetic localization algorithms in different scale scenario.}
     \label{fig:cmpplotcdfareascale}
     \end{figure}
   
   
   Fig.~\ref{fig:cmpplotcdfareascale} shows the cumulative error distributions of different localization algorithms in different scale scenarios. To make the comparison result more convincing, we construct the neural network of stacking the multi-scale TCN before the TCN (MSTT). The proposed MSTL has a better localization accuracy than other magnetic localization algorithms. In different scale scenarios, the proposed MSTL can keep the errors of more than $90\%$ less than 2 m. However, other localization algorithms the percentage of errors less than 2 m decreases as the scenario scales up. TABLE \ref{tab:sceresult} presents the localization results more detailedly. The average errors of the proposed MSTL in the sceID S, M, L are 0.75 m, 1.04m, and 1.03m. In comparison with DTW, LSTM and TCN, the MSTT algorithm has no better localization accuracy. The results also reveals that the stacking of TCNs alone does not improve the localization accuracy, rather the stacking of TCN and LSTM does. 
   \begin{table}[!t]
   \centering
   \begin{threeparttable}
   \caption{Comparison of localization error regarding different localization algorithms and scenarios}
   \label{tab:sceresult}
   \small 
   \centering
   \begin{tabular}{lccccccc}
   \toprule
   &\multicolumn{2}{c}{\textbf{Error of SceID S}}&\multicolumn{2}{c}{\textbf{Error of SceID M}}&\multicolumn{2}{c}{\textbf{Error of SceID L}}\\
   \cmidrule(l){2-3}
   \cmidrule(l){4-5}
   \cmidrule(l){6-7}
   \multirow{-2}{*}{\textbf{Algorithm}}&\textbf{Mean}&\textbf{SD}&\textbf{Mean}&\textbf{SD}&\textbf{Mean}&\textbf{SD}\\
   \midrule
   DTW &0.81&0.730&1.43&1.913&1.76&2.667\\
   LSTM&1.42&0.643&1.57&0.868&2.03&1.913\\
   TCN &1.01&0.606&1.48&1.000&1.44&2.067\\
   MSTT&0.84&0.568&1.65&1.263&3.23&7.992\\
   MSTL&\textbf{0.75}&\textbf{0.465}&\textbf{1.04}&\textbf{0.865}&\textbf{1.03}&\textbf{1.111}\\
   \bottomrule
   \end{tabular}
   \begin{tablenotes}
   \item[Mean] The mean of the localization errors
   \item[SD] The standard deviation of the localization errors
   \end{tablenotes}
   \end{threeparttable}
   \vskip -0.6cm
   \end{table}

   \begin{table}[!t]
   \centering
   \begin{threeparttable}
   \caption{Comparison of localization error regarding different localization algorithms and test speeds}
   \label{tab:speedresult}
   \small 
   \centering
   \begin{tabular}{lcccccccccc}
   \toprule
   &\multicolumn{2}{c}{\textbf{DTW}}&\multicolumn{2}{c}{\textbf{LSTM}}&\multicolumn{2}{c}{\textbf{TCN}}&\multicolumn{2}{c}{\textbf{MSTT}}&\multicolumn{2}{c}{\textbf{MSTL}}\\
   \cmidrule(l){2-3}
   \cmidrule(l){4-5}
   \cmidrule(l){6-7}
   \cmidrule(l){8-9}
   \cmidrule(l){10-11}
   \multirow{-2}{*}{\textbf{SP}}&\textbf{Mean}&\textbf{SD}&\textbf{Mean}&\textbf{SD}&\textbf{Mean}&\textbf{SD}&\textbf{Mean}&\textbf{SD}&\textbf{Mean}&\textbf{SD}\\
   \midrule
   \rowcolor{mygray1}1/8& 0.84&0.748 &1.37&0.750&1.01&0.630&0.87&0.507&0.60&0.412\\ 
                     1/7& 0.84&0.747 &1.37&0.750&1.01&0.630&0.86&0.507&0.60&0.412\\ 
                     1/6& 0.84&0.747 &1.37&0.749&1.01&0.630&0.86&0.507&0.60&0.412\\ 
                     1/5& 0.84&0.746 &1.37&0.749&1.01&0.630&0.86&0.506&0.60&0.411\\ 
   \rowcolor{mygray1}1/4& 0.83&0.744 &1.37&0.747&1.01&0.629&0.86&0.505&0.60&0.411\\ 
                     1/3& 0.83&0.742 &1.36&0.745&1.01&0.627&0.86&0.504&0.60&0.410\\ 
   \rowcolor{mygray1}1/2& 0.83&0.738 &1.36&0.740&1.00&0.625&0.86&0.502&0.59&0.410\\ 
     \rowcolor{mygray1}1& 0.81&0.730 &1.34&0.736&1.00&0.619&0.86&0.507&0.58&0.410\\ 
     \rowcolor{mygray1}2& 1.51&1.289 &1.47&0.755&1.65&1.238&1.36&0.881&0.63&0.394\\ 
                       3& 3.96&4.316 &1.50&0.947&2.39&1.754&1.81&1.542&0.64&0.470\\ 
     \rowcolor{mygray1}4&14.34&12.301&1.62&0.905&2.83&2.193&1.72&1.394&0.78&0.536\\ 
                       5&29.04&10.861&1.54&0.830&5.01&3.496&4.41&4.017&1.29&0.944\\ 
                       6&35.77&7.490 &1.50&0.742&4.33&4.212&3.34&4.121&0.97&0.785\\ 
                       7&40.00&5.762 &1.55&1.001&3.70&2.738&1.91&2.018&0.70&0.585\\ 
     \rowcolor{mygray1}8&43.44&6.011 &1.55&0.693&4.43&2.969&3.16&2.132&1.25&0.894\\ 
   \bottomrule
   \end{tabular}
   \begin{tablenotes}
   \item[SP] The multiple of sampling speed
   \end{tablenotes}
   \end{threeparttable}
   \vskip -0.3cm
   \end{table}
   
   TABLE \ref{tab:speedresult} presents the comparison results of the proposed localization algorithm, DTW, LSTM, TCN, and MSTT at different user's moving speeds. The magnetic data at 2$\sim$8 times moving speeds are generated by extracting data with interval of 2$\sim$8 from the sampling magnetic signals of sceID S. The magnetic data with 1/8$\sim$1/2 moving speeds are generated by 2$\sim$8 times linear interpolation of the sampling magnetic signals. The magnetic data with 1/8, 1/4, 1/2, 1, 2, 4, 8 times speeds are used to train localization model. All the magnetic data are used to evaluate the performances of the different localization algorithms. The propsed algorithm has a better localization accuracy compared with other localization algorithms. The experiments with 1/8$\sim$1/2 moving speeds have similar localization results, since the magnetic features interpolated linearly are similar to partial sampling magnetic features. However, the experiments with 2$\sim$8 times moving speeds have different localization results. The accuracy of DTW localization algorithm decreases as the moving speed increases. The DTW localization exists the proble of mismatching at different speeds. For the TCN and MSTT algorithms, the accuracy at the speed of training localization models is better than others. The generalization abilities of TCN and MSTT are weak. The LSTM and MSTL do not have the aforementioned problems of DTW, TCN and MSTT. The MSTL has the better localization performance than than the LSTM.
   
   \begin{figure}[!t]
   \centering
   \vspace{-0.2cm}
   \subfloat[LSTM Localization]{\includegraphics[height=6cm]{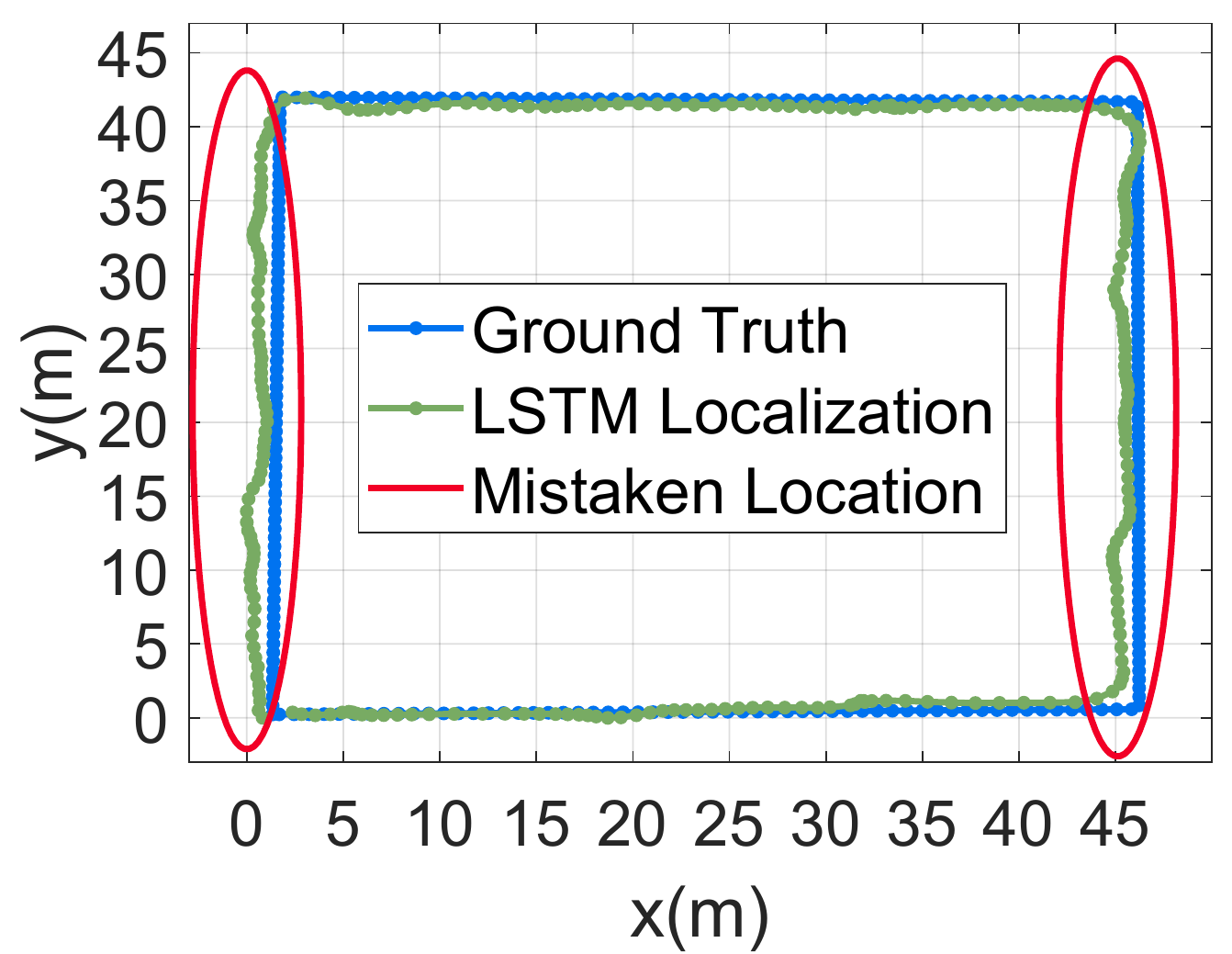}%
   \label{fig:trajectorya}}
   \hfil
   \subfloat[TCN Localization]{\includegraphics[height=6cm]{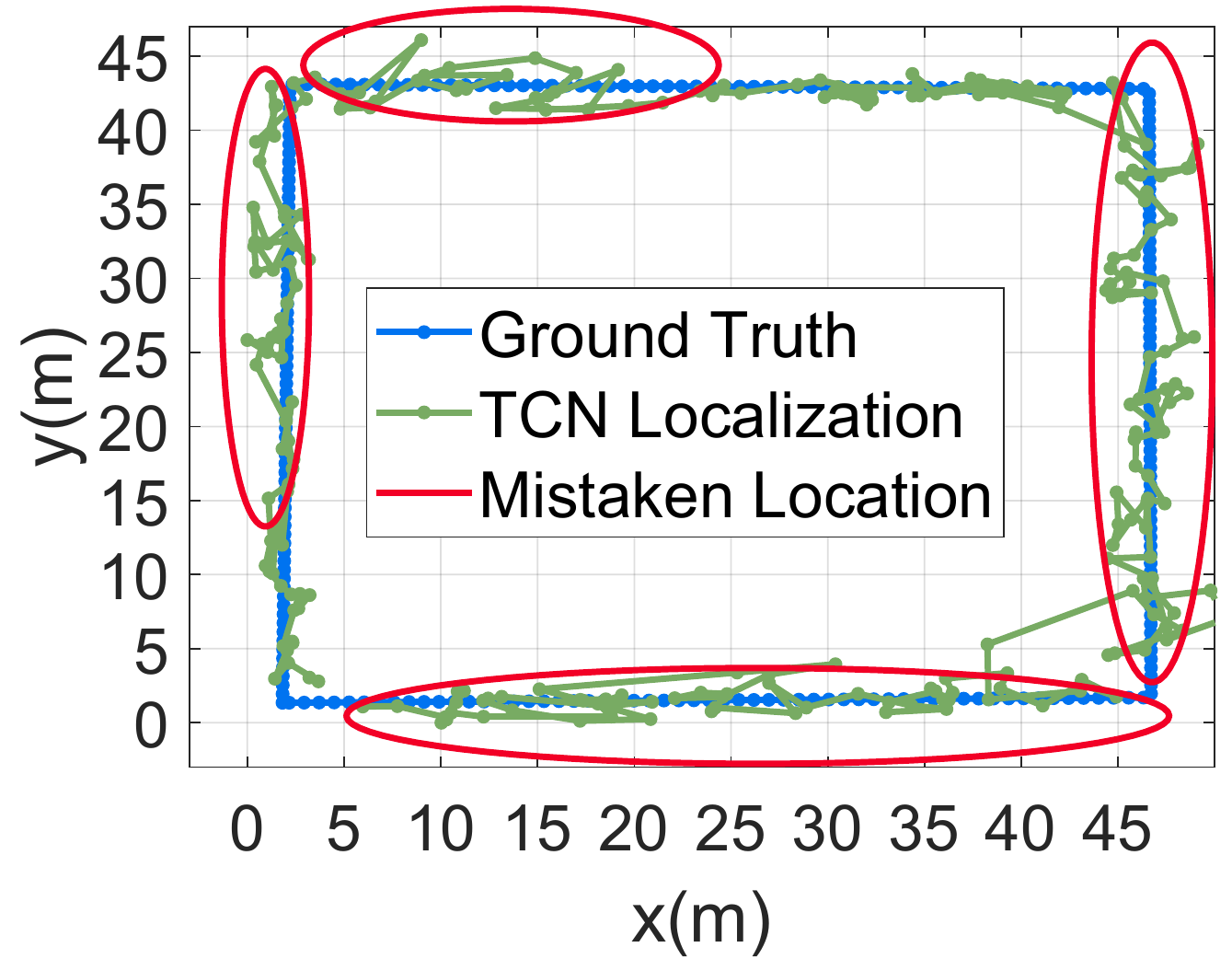}%
   \label{fig:trajectoryb}}
   \hfil
   \subfloat[MSTL Localization]{\includegraphics[height=6cm]{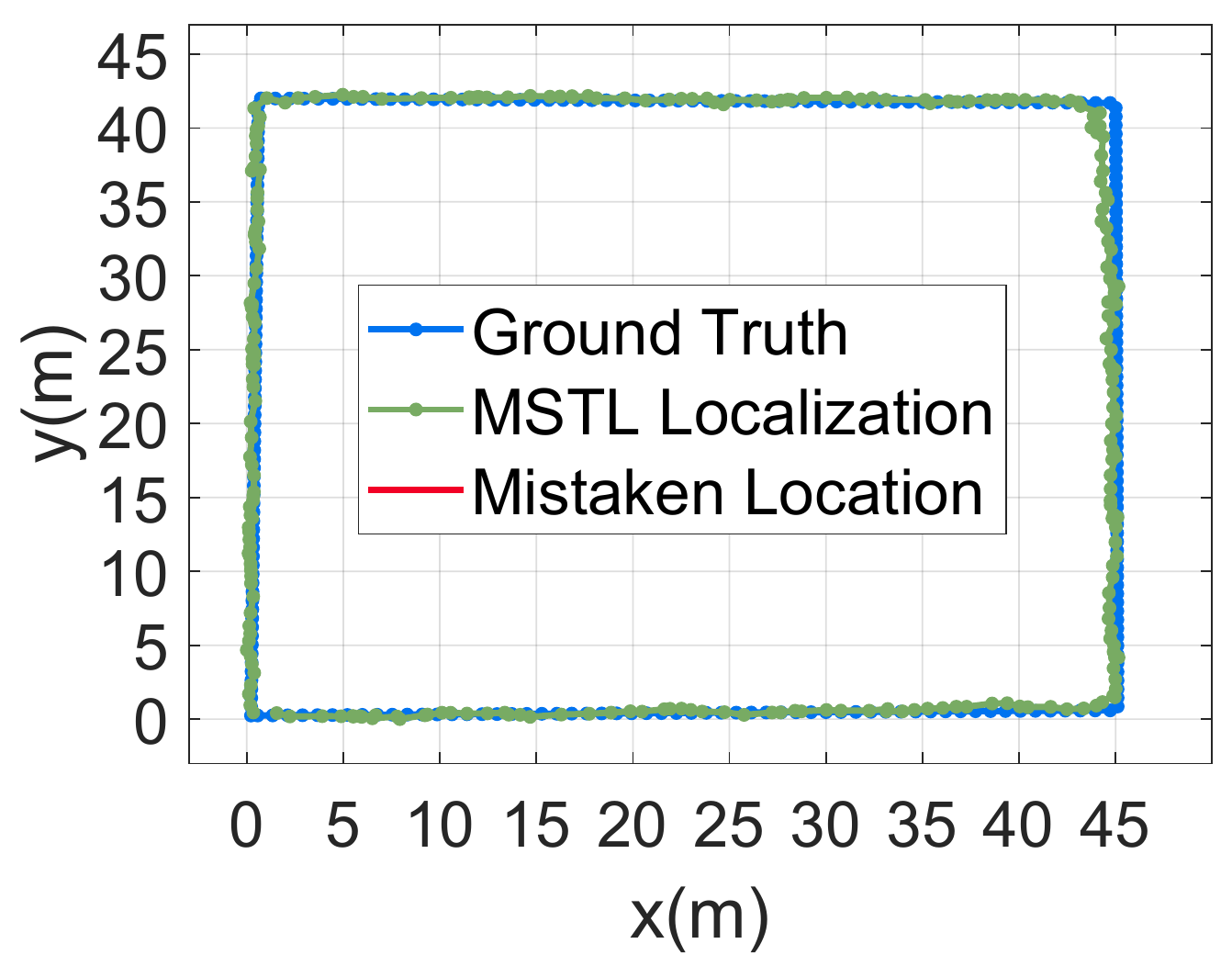}%
   \label{fig:trajectoryc}}
   \hfil
   \subfloat[MSTT Localization]{\includegraphics[height=6cm]{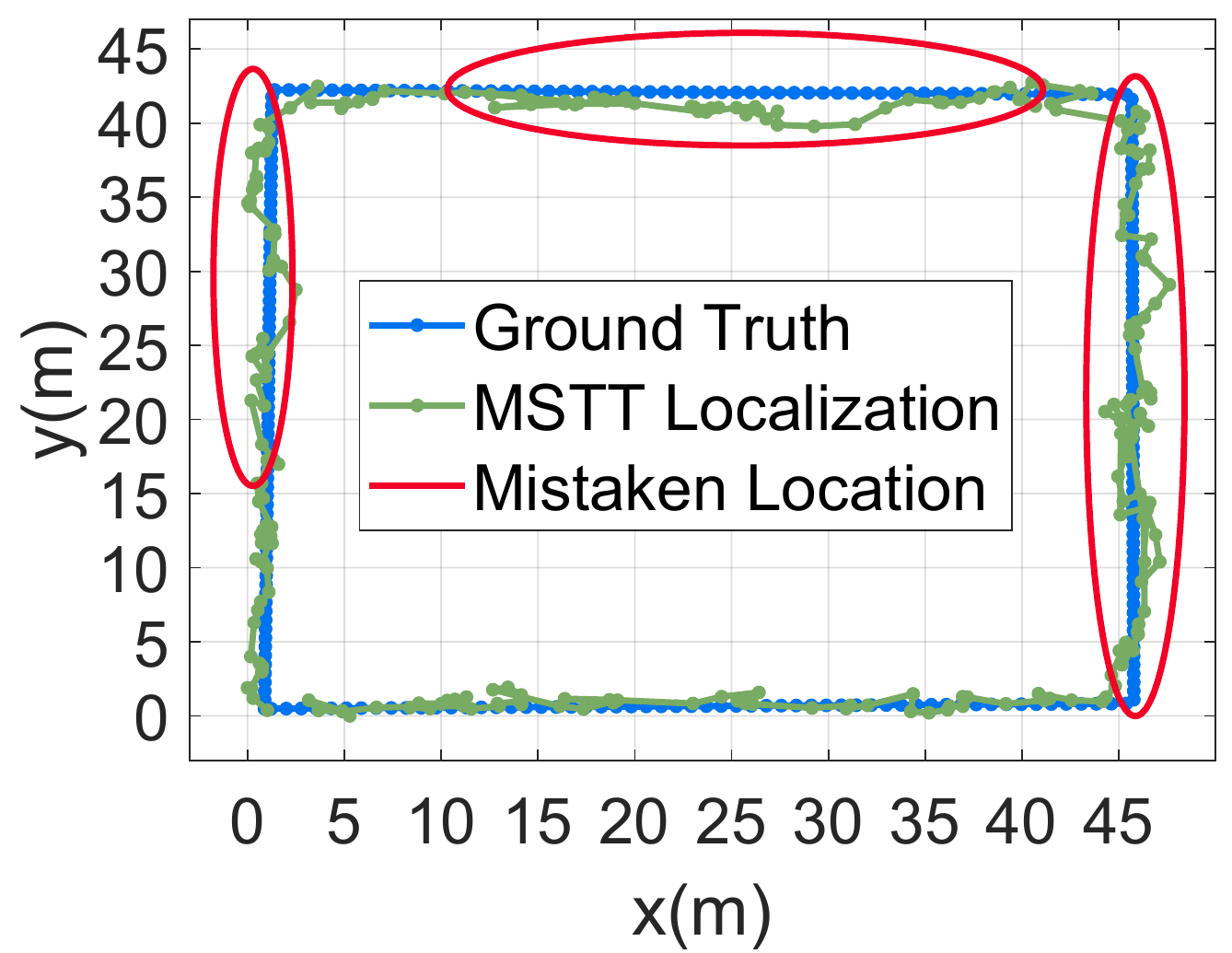}%
   \label{fig:trajectoryd}}
   \caption{The trajectories of different localization algorithms.}
   \label{fig:trajectory}
   \end{figure}
   Fig.~\ref{fig:trajectory} depicts the trajectories of different localization algorithms at 3 times moving speed. The mistaken locations of the TCN algorithm are the discrete localization points, however, those of the LSTM algorithm are the continuous localization points. Since the TCN uses the causal convolutions to simulate the temporal context, the generalization abilities of TCN is weaker than that of LSTM. If the user's temporal characteristics are not fully consistent with the TCN model, the localization results will be inaccurate. Therefore, the mistaken locations of the TCN algorithm are the discrete localization points. The MSTT has the similar problem. Since the feature dimension of LSTM is fixed 3. A continuous segment of low-discernibility geomagnetic data will be continuously used many times for LSTM localization. Therefore, the mistaken locations of the LSTM algorithm are the continuous localization points. However, the proposed algorithm increases the feature dimension of LSTM by invoking multi-scale TCN. Therefore, the proposed LSTMs-based localization can provide a better localization performance at different user's moving speeds.

   
   \section{Conclusion}\label{sec:con}
   
   This paper proposes a novel multi-scale TCN and LSTM-based indoor magnetic localization algorithm for smartphones. We have preprocessed the raw geomagnetic signals into a time-series form, thus enhancing the discernibility of geomagnetic signals. we have invoked the TCN to expand the feature dimensions on the basis of keeping the time-series characteristics of LSTM model. To address the problem of inconsistent time-series speed between localization model and mobile users, we have constructed a multi-scale time-series layer with multiple TCNs of different dilation factors. We have proposed a stacking framework of multi-scale TCN and LSTM for indoor magnetic localization. We have evaluated our localization algorithm under a variety of experimental scenarios and demonstrated the effectiveness of our proposed algorithm on indoor localization problems.

   \ifCLASSOPTIONcaptionsoff
     \newpage
   \fi

   
   
   \bibliographystyle{IEEEtran}
   %

\end{document}